\documentclass[amssymb]{elsart}
\usepackage{graphicx,here} 
\newcommand{\beq}{\begin{equation}}
\newcommand{\eeq}{\end{equation}}
\newcommand{\ba}{\begin{array}}
\newcommand{\ea}{\end{array}}
\newcommand{\lsim}   {\mathrel{\mathop{\kern 0pt \rlap
  {\raise.2ex\hbox{$<$}}}
  \lower.9ex\hbox{\kern-.190em $\sim$}}}
\newcommand{\gsim}   {\mathrel{\mathop{\kern 0pt \rlap
  {\raise.2ex\hbox{$>$}}}
\lower.9ex\hbox{\kern-.190em $\sim$}}}
\begin{document}
\begin{frontmatter}
\title{Super Heavy Dark Matter and UHECR Anisotropy at Low Energy}
\author{Roberto Aloisio$^a$\thanksref{corr1}, Francesco Tortorici$^{a,b}$}
\address[lngs]{INFN, Laboratori Nazionali del Gran Sasso, SS 17bis, Assergi (Italy)}
\thanks[corr1]{E-mails:\\roberto.aloisio@lngs.infn.it\\francesco.tortorici@lngs.infn.it}
\address[AQ]{Dipartimento di Fisica, Universit\`a di L'Aquila, \\
via Vetoio I-67010, L'Aquila, Italy}

\begin{abstract}
Super Heavy quasi-stable particles are naturally produced in the early universe and could represent a substantial fraction of the Dark Matter: the so-called Super Heavy Dark Matter (SHDM). The decay of SHDM represents also a possible source of Ultra High Energy Cosmic Rays (UHECR), with a reliably calculated spectrum of the particles produced in the decay $(\propto E^{-1.9})$. The SHDM model for the production of UHECR can explain quantitatively only the excess of UHE events observed by AGASA. In the case of an observed spectrum not showing the AGASA excess the SHDM model can provide only a {\it subdominant} contribution to the UHECR flux. We discuss here the basic features of SHDM for the production of a {\it subdominant} UHECR flux, we refer our study to the possible signatures of the model at the Auger observatory discussing in particular the expected chemical composition and anisotropy.
\end{abstract}

\end{frontmatter}

\section{Introduction}  
\label{sec:introduction}

Ultra High Energy Cosmic Rays (UHECR) are the most energetic particles known in nature with energies exceeding $10^{20}$ eV, the observation of particles with such high energies raises many interesting questions about their origin and composition that involve both astrophysics and particle physics. 

Soon after the discovery of the Cosmic Microwave Background (CMB) radiation it was shown that the flux of UHE protons should be characterized by a sharp steepening at energy $\sim 5\times 10^{19}$~eV, due to the photo-pion production process on the CMB radiation field \cite{GZK}. This effect is the well known Greisen-Zatsepin-Kuzmin (GZK) cut-off.  After a few decades of observations the detection of the GZK steepening of the CR flux is one of the major open problems in CR physics with experimental data still not conclusive. The AGASA experiment \cite{AGASA} observed 11 events with energy $E>10^{20}$ eV in contrast with the expected depletion due to the GZK effect; on the other hand the HiRes experiment \cite{HiRes} observed an UHECR flux in agreement with the GZK cutoff \cite{HiResGZK}. The discrepancy between these two experimental results has been widely discussed and recently it has been shown that the GZK feature cannot be accurately determined with the small sample of events collected by AGASA and HiRes and that the discrepancy between the two experiments has a low statistical significance (at a $2\sigma$ level) \cite{BlasiDeMarco}. Moreover, AGASA and HiRes are based on different experimental techniques: a ground array the former and a fluorescence detector the latter. Taking this difference into account, the differing results of AGASA and HiRes can be interpreted assuming the presence of a $30\%$ systematic error in the relative energy determination. As shown by \cite{Dip,BlasiDeMarco}, correcting for these systematics the observations of AGASA and HiRes are brought into agreement at energies $E<10^{20}$ eV and the discrepancy at the highest energies is softened. Recently, the Pierre Auger Observatory \cite{Auger}, under completion in Argentina, published its first observations of the UHECR flux combining the ground based and fluorescence detection techniques. This preliminary result seems to confirm the presence of the GZK steepening \cite{AugerFlux}.

The presence of an excess of very high energy events, as claimed by AGASA, inspired the introduction of several exotic models for the production of UHECR. These models, collectively called top-down, reproduce the excess of AGASA and give also an explanation for the lack of any clear astrophysical counterpart to the highest energy events observed. Indeed, at energies $E>5\times 10^{19}$ eV the proton attenuation lenght is only about $20\div 30$ Mpc and an astrophysical source at this distance should have been seen at least in different frequency ranges. This evidence could imply that particles with energy $E\sim 10^{20}$ eV have a different (top-down) origin respect to those at lower energies. Many different ideas have been proposed among top-down models: strongly interacting neutrinos \cite{nu} and new light hadrons \cite{gluino} as unabsorbed signal carriers, $Z$-bursts \cite{Z}, Lorentz-invariance violation \cite{Lorentz}, Topological Defects (TD) (see \cite{BBV,TD} for a review), and Superheavy Dark Matter (SHDM) (see \cite{SHDM} for a review).

The two models based on SHDM and TD have common features: in both cases UHECR are produced in the decay (SHDM) or annihilation (TD) of super-heavy particles, with a typical mass $M_{X}>10^{13}$ GeV, that we will call collectively X-particles. As already discussed in \cite{ABK}, TD are distributed over cosmological distances therefore give only a marginal contribution to the  UHECR flux. We will not discuss this case here, concentrating our attention on the SHDM hypothesis. The possible existence of super heavy relic particles is an interesting conclusion of modern cosmology, being first suggested in connection with UHECR production \cite{BKV,KR} and later developed as a suitable candidate for Dark Matter (DM). 

Two main problems should be addressed in the discussion of SHDM models: how particles with very high mass ($M_{X}>10^{13}$ GeV) can be quasi-stable, with a lifetime much longer than the age of the universe $t_0$, and how their abundance can be dominant in the universe today. The stability of SHDM can be achieved assuming the existence of a discrete gauge simmetry that protects the particle from decaying, in the same way as neutralino stability through R-parity in Super Symmetry (SUSY). This discrete simmetry can be weakly broken, assuring a lifetime $\tau_X>t_0$, through wormhole \cite{BKV} or instanton \cite{KR} effects, an example of a particle with a lifetime exceeding the age of the universe can be found in \cite{crypton}. The abundance of SHDM can easily be dominant in the universe today, with a SHDM density $\Omega_{SHDM}\sim \Omega_{DM}$. This effect can be obtained by gravitational production that resembles the production of density fluctuations during inflation \cite{ZS}.

The top-down hypothesis and, in particular, the UHECR production through SHDM decay can account only for the highest energy part of the observed spectrum, as in the case of AGASA excess at energies 
$E>5\times 10^{19}$ eV \cite{ABK}. However, UHECR observations cannot exclude SHDM as explanation of the DM problem, assuming that SHDM is gravitationally produced than the X particle mass and density are unambiguously fixed, the only free parameter left to fit UHECR observations is the life-time $\tau_X$. The observation of the AGASA excess fixes this value as $\tau_X\simeq 10^{20}$ y \cite{ABK,ABK2}. From the HiRes, Yakutsk and Auger data, that are compatible with the GZK steepening, follows only a lower bound on the value of $\tau_X$. Within this lower bound it is still possible to test the SHDM hypothesis being connected with a {\it subdominant} contribution to the observed UHECR flux. In the present paper we will discuss such a possibility referring in particular to the observations of chemical composition and anisotropy of the Auger observatory.

The UHECR production through the decay of SHDM shows three basic signatures that can be used to test the model: 
\begin{itemize}
\item{SHDM particles, as any DM candidate, are clustered by gravitational interaction and accumulated in the halo of our galaxy with a typical over-density of $\delta\sim 2\times 10^{5}$ \cite{BKV}. Hence the UHECR spectrum from SHDM can overcome the constrain of the GZK steepening \cite{BKV}.}
\item{In the decay of X-particles pions are extensively produced \cite{ABK2}, therefore UHE neutrinos 
and photons are the dominant component of the primary flux \cite{ABK2}.} 
\item{The non-central position of the Sun in the galactic halo results in an anisotropic flux of UHECR observed on earth \cite{ABK2,DT}.} 
\end{itemize}

The photons (and neutrinos) dominated flux expected from SHDM is one of the most striking predictions of the model. Recently, the expected SHDM photon flux at energies $E>10^{19}$ eV was compared with the photon fraction in the AGASA events \cite{ABK2}: this fraction was determined by different groups with different results, being around $50\%$ in \cite{Rubtsov} and higher $67\%$ in \cite{Risse}. Another important analysis on the photon content in UHECR events was recently published by the Auger collaboration \cite{AugerGamma}. From this analysis, performed at the highest ($E \gsim 10^{19}$ eV) energies, follows a stringent limit on the photon fraction that is around $2\%$ at $10^{19}$ eV and below $40\%$ at $10^{20}$ eV 
\cite{AugerGamma}. As recently discussed in \cite{SemikozICRC07}, these limits do not exclude, but disfavor, the SHDM hypothesis for the explanation of the observed UHECR spectrum, in the next section we will come back to the expected photon fraction in SHDM models. 

The second most important signature of UHECR production by SHDM is the peculiar anisotropy expected. This anisotropy in the flux is guaranteed by our position respect to the Galactic Center (GC). In particular, the distance between the Sun and the outer boundary of the galaxy is larger in the GC direction respect to the anti-center direction. In order to evaluate the expected anisotropy it is necessary to assume a particular distribution of SHDM in the galaxy: in this respect numerical simulations of the galactic DM distribution show an increase of the DM density towards the GC as $\propto r^{-1}$ \cite{NFW} or $\propto r^{-1.5}$ \cite{Moore}, this further enhancing the expected anisotropy. 

Already several authors have considered anisotropy computations, with reliable predictions of the expected signal in the highest energy part of the spectrum \cite{BBV,ABK2,DT}, where the SHDM contribution can be dominant. The expected anisotropy was compared with the existing UHECR data taken in the northern hemisphere, this comparison reveals no contradiction between data and the predictions of the model \cite{BM},\cite{TW}. On the other hand, the detectors located in the southern hemisphere are able to observe the GC and are much more sensitive to the expected anisotropy: in this respect the data of the old Sugar detector, in Australia, are  only marginally consistent with the predictions of the SHDM model \cite{KS}.

The Pierre Auger Observatory, due to its location in the southern emisphere and its large exposure, has a big potential for the detection of UHECR from the GC and the related anisotropy. In the present paper we will discuss such a potential, with particular attention to the lower energy region accessible to the detector, namely the range $3\times 10^{18}\div 3\times 10^{19}$ eV.  At these energies the SHDM contribution to the flux, being {\it subdominant}, can be extracted from the data only using its peculiar anisotropy pattern. Although {\it subdominant}, UHE particles coming from the decay of SHDM are mainly photons that do not suffer any isotropization process by magnetic fields, while $3\times 10^{18}\div 3\times 10^{19}$ eV particles produced by astrophysical sources, being charged, are deflected by the galactic magnetic field \cite{BlasiGalB} showing a substantially isotropic flux. In this sense, the low energy data of the Auger detector, with high statistics, can be used to shad light on the SHDM hypothesis, being any observed anisotropy directly connected only with SHDM.

In order to compute the extragalactic UHECR flux we have used the Dip model, recently suggested to explain the behavior of UHECR at energies $E>10^{18}$ eV \cite{Dip}. In this model the observed UHECR flux in the energy range $10^{18} {\rm eV}~\div~8\times 10^{19} {\rm eV}$ is explained as the effect of the combination of the adiabatic and pair-production energy losses suffered by protons interacting with the CMB field.  This interpretation of the observed spectra \footnote{The observed UHECR spectrum has also an alternative interpretation based on the so-called mixed composition scenario \cite{Allard}. In this model UHECR with energies $E>10^{19}$ eV are constituted by a mixture of extragalactic protons and heavy nuclei, with a flat injection spectrum $\gamma<2.3$, while, at lower energies $E<10^{19}$ eV UHECR have still a galactic origin \cite{Allard}. From the point of view of the SHDM induced anisotropy the "Dip" and "mixed composition" scenarios show similar features.} has important consequences mainly from the point of view of the origin and chemical composition of UHECR, implying that, already at energies $E>10^{18}$ eV, UHECR have an extragalactic origin and are mainly constituted by protons, with a steep injection spectrum $\propto E^{-\gamma}$ $\gamma>2.5$.

The paper is organized as follows: in section \ref{sec:flux} we will introduce the computation of the expected flux from SHDM combining it with the extragalactic proton flux as in the Dip model, discussing in particular the expected chemical composition at the Auger observatory. In section \ref{sec:ani} we will calculate the anisotropy of UHECR, referring our computation to the case of the Auger detector. Finally in section \ref{sec:concl} we will conclude discussing our results.

\section{Flux and chemical composition}
\label{sec:flux}

The first step to determine the UHECR flux produced in the decay of $X$ particles is the computation of the particles spectra. As it was already discussed in \cite{ABK}, this evaluation is particularly important because it represents a direct signature of the production mechanism that, in principle, can be detected experimentally. As discussed in the introduction, the mass of the decaying particle, $M_X$, that represents the total CMS energy $\sqrt{s}$ in the decay, is in the range $10^{13} \div 10^{16}$ GeV. From the point of view of elementary particle physics the X-particle decay proceeds in a way similar to the $e^{+}e^{-}$ annihilation into hadrons, two or more off mass shell partons are produced initiating QCD cascades.

The determination of the particles energy spectra produced in the X-particle decay requires the extension of the existing QCD calculations for parton cascade from the TeV scale up to the $M_X$ energy scale. The first calculations proposed in literature used an analytical approach based on the limiting spectrum approximation \cite{limiting} or an extension of the HERWIG \cite{herwig} QCD Monte Carlo (MC) 
\cite{sarkar}. More recently several papers appeared discussing the computation scheme involved in the SHDM decay. These schemes are based on two different approaches: the DGLAP evolution equations \cite{ABK,BD,ST} and a SUSY-QCD MC approach generalized to the high energies involved in the decay \cite{BK,ABK}. These computations predict quite accurately the secondary spectra from the X-particle decay
and agree well with each other. Their most important outcome, concerning UHECR physics, is a flat energy spectrum of the produced particles with a behavior that can be approximated as $dE/E^{1.9}$ 
\cite{ABK}.

As already discussed in the introduction, an important signature of the SHDM model is represented by the 
peculiar UHECR composition, with a photon dominated flux. The exact determination of the photon fraction in the expected spectra is another outcome of the computation schemes discussed above, the typical photon/nucleon fraction expected in the cascade is $r_{\gamma/N}\simeq 2\div 3$ \cite{ABK}, being only weekly dependent on the energy.

The UHECR emissivity produced by the decays of SHDM in the halo of the galaxy can be simply written 
as:
\begin{equation}
I_{SHDM}(E) = \frac{1}{M_X \tau_X} D(E) n_X(R)
\label{emiss}
\end{equation}
where $D(E)$ is the energy spectrum (fragmentation function) of particles (photons and protons \footnote{In the present paper we will not discuss neutrinos from SHDM, that is the subject of a forthcoming paper 
\cite{AB08}}) produced by the SHDM decay, and $n_{X} (R)$ is the density of X-particles in the galaxy as function of the distance $R$ from the GC. In the present work we will assume that X-particles contribute to a substantial fraction of the galactic DM: therefore we will use the DM density as obtained by several numerical simulations. In particular we will use the Navarro-Frank-White (NFW) \cite{NFW} and Moore \cite{Moore} DM density profiles
\begin{equation}
n_{X} (R) = \frac{n_0}{(R/R_s)^\alpha (1+R/R_s)^{3-\alpha}}
\label{profile}
\end{equation}
with $\alpha=1,3/2$ for the NFW and Moore case respectively, and $R_s=45$ Kpc as obtained in \cite{BM}. 

\begin{figure}[ht]
\includegraphics[width=0.52\textwidth]{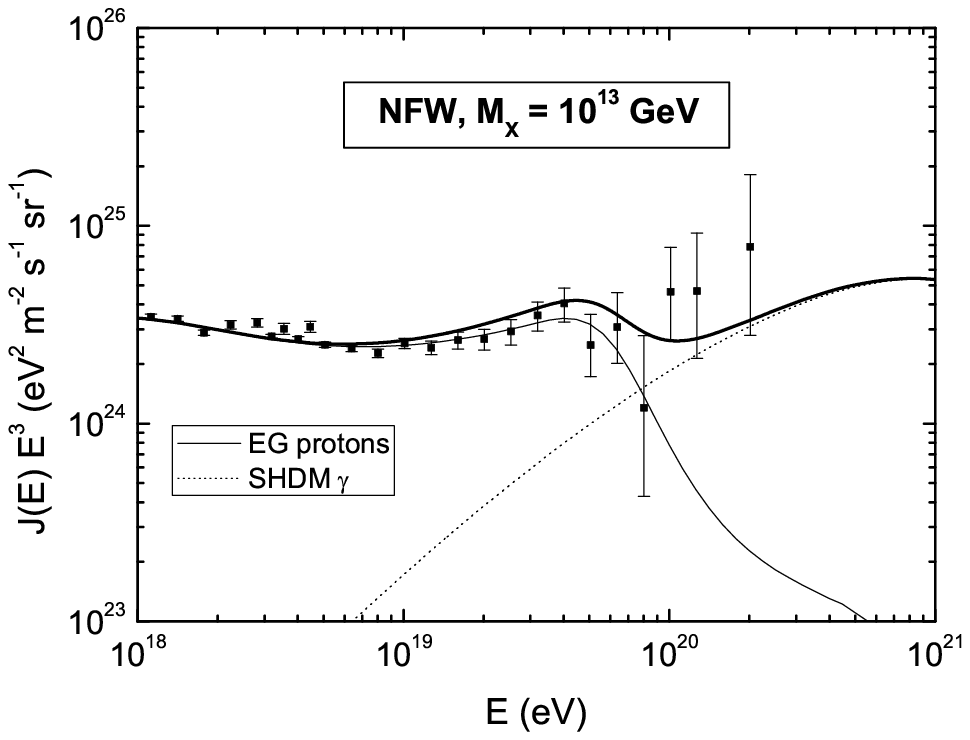}
\includegraphics[width=0.52\textwidth]{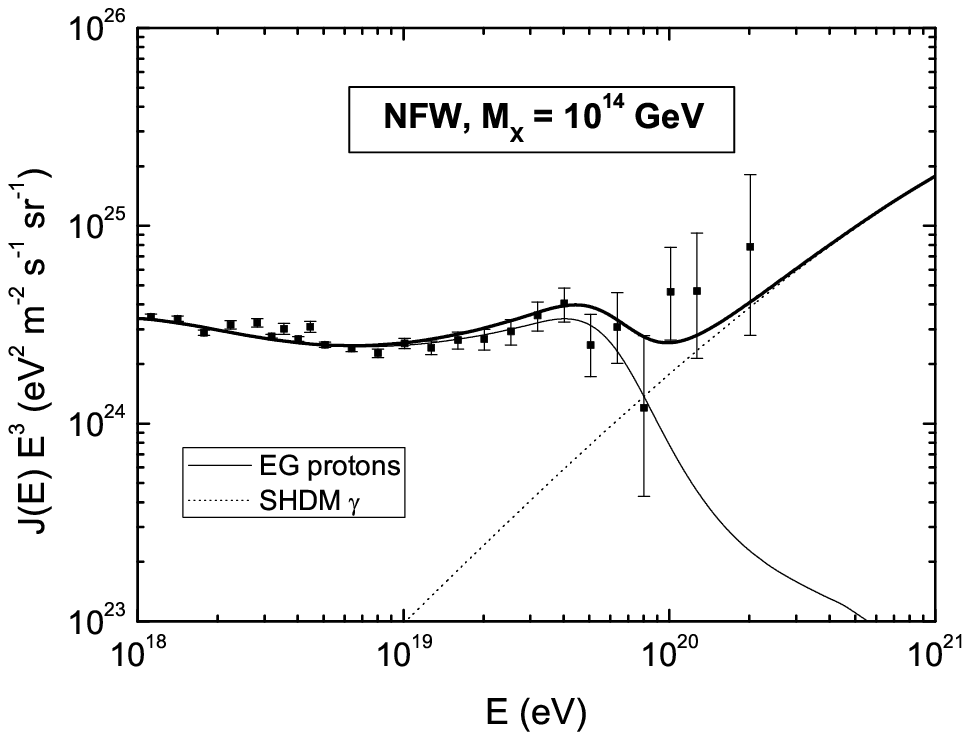}
\caption{The calculated spectrum of UHECRs from SHDM (dotted curve) in comparison with the AGASA data ($M_X=10^{13}$ GeV left panel, $M_X=10^{14}$ GeV right panel). The spectrum from SHDM decays can explain only the highest energy events (AGASA excess). The continuos curve gives the
spectrum of extragalactic protons (assuming the Dip model \cite{Dip}). The sum of these two spectra is shown by thick continuos curve.}
\label{fig1}
\end{figure} 

Once determined the UHECR emissivity the expected flux can be simply computed as the line of sight integral of the emissivity (\ref{emiss}). Using the spherical symmetry of the DM density profile $n_X(R)$ we can identify the single line of sight simply by one angular variable $\vartheta$, that represents the angle formed by the axis Sun-GC and the line of sight itself. Therefore, the UHECR flux from the SHDM decay will be 

\begin{equation}
J_{SHDM}(E,\vartheta)=\frac{1}{4\pi M_X \tau_X} D(E) \int_0^{r_{max}(\vartheta)} dr n_{X} (R(r))
\label{flux1}
\end{equation}
where $r$ is the line of sight coordinate and $r_{max}(\vartheta)$ is the distance to the boundary of the galactic halo in the $\vartheta$ direction
\begin{equation}
r_{max}(\vartheta)=R_\odot \cos(\vartheta) + \sqrt{R_h^2 + R_\odot^2 \sin^2\vartheta}
\label{rmax}
\end{equation}
where $R_h=100$ Kpc is the radius of the galactic halo and $R_\odot=8.5$ Kpc is the distance Sun-GC.

Let us now introduce the extragalactic (EG) component of the UHECR flux. As discussed in the introduction, we will use the Dip model \cite{Dip} assuming an homogeneous distribution of sources and a proton dominated injection at the source. Under these hypothesis the UHECR flux can be computed as
\begin{equation}
J_{EG} (E)=\frac{c}{4\pi} \int_0 dz \left | \frac{dt}{dz} \right | Q_{inj}(E_g(E,z)) \frac{dE_g(E,z)}{dE}
\label{universal}
\end{equation}
where: $dt/dz=(H_0(1+z)\sqrt{\Omega_m (1+z)^3 + \Omega_\Lambda})^{-1}$ represents the standard cosmology ($\Omega_m=0.27$ and $\Omega_\Lambda=0.73$ \cite{WMAP}); $Q_{inj}(E_g) \propto E_g^{-\gamma}$ is the number of particles injected at the source per unit volume and energy, being the best fit value in the Dip model $\gamma_g=2.7$ \cite{Dip}; the generation energy $E_g(E,z)$ is determined solving the losses equation $dE_g/dt = b(E_g,t)$ with the initial condition $E_g(t=t_0)=E$ \cite{Dip}; the quantity $dE_g/dE$ can be computed following the recipe given in \cite{Dip,BG}.

The total UHECR flux will be the sum of the two contributions: galactic (from SHDM) and extragalactic (from homogeneous distributed sources)
\begin{equation}
J_{UHECR}(E,\vartheta) = J_{EG}(E) + J_{SHDM}(E,\vartheta)
\label{total1}
\end{equation}

\begin{figure}[ht]
\includegraphics[width=0.52\textwidth]{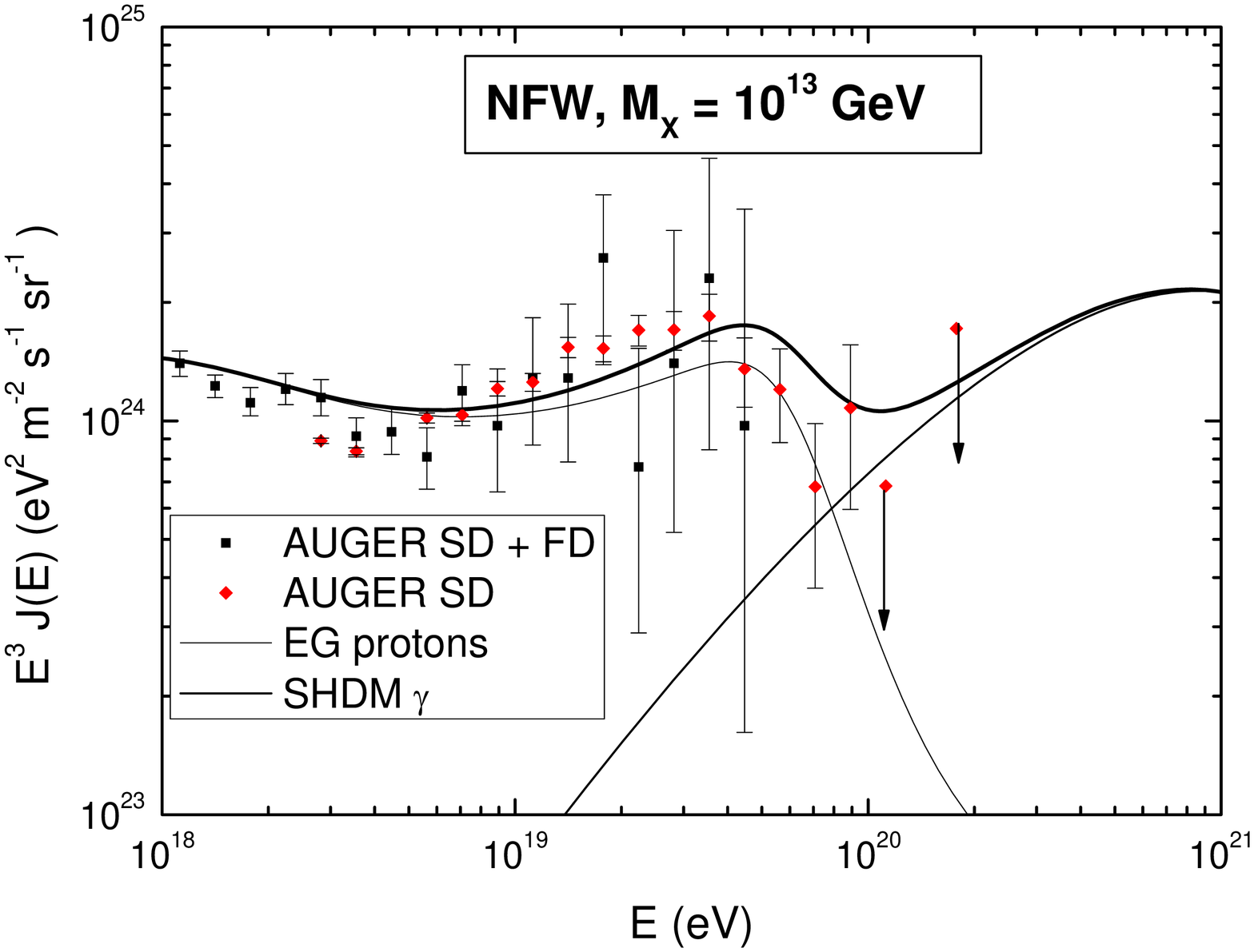}
\includegraphics[width=0.52\textwidth]{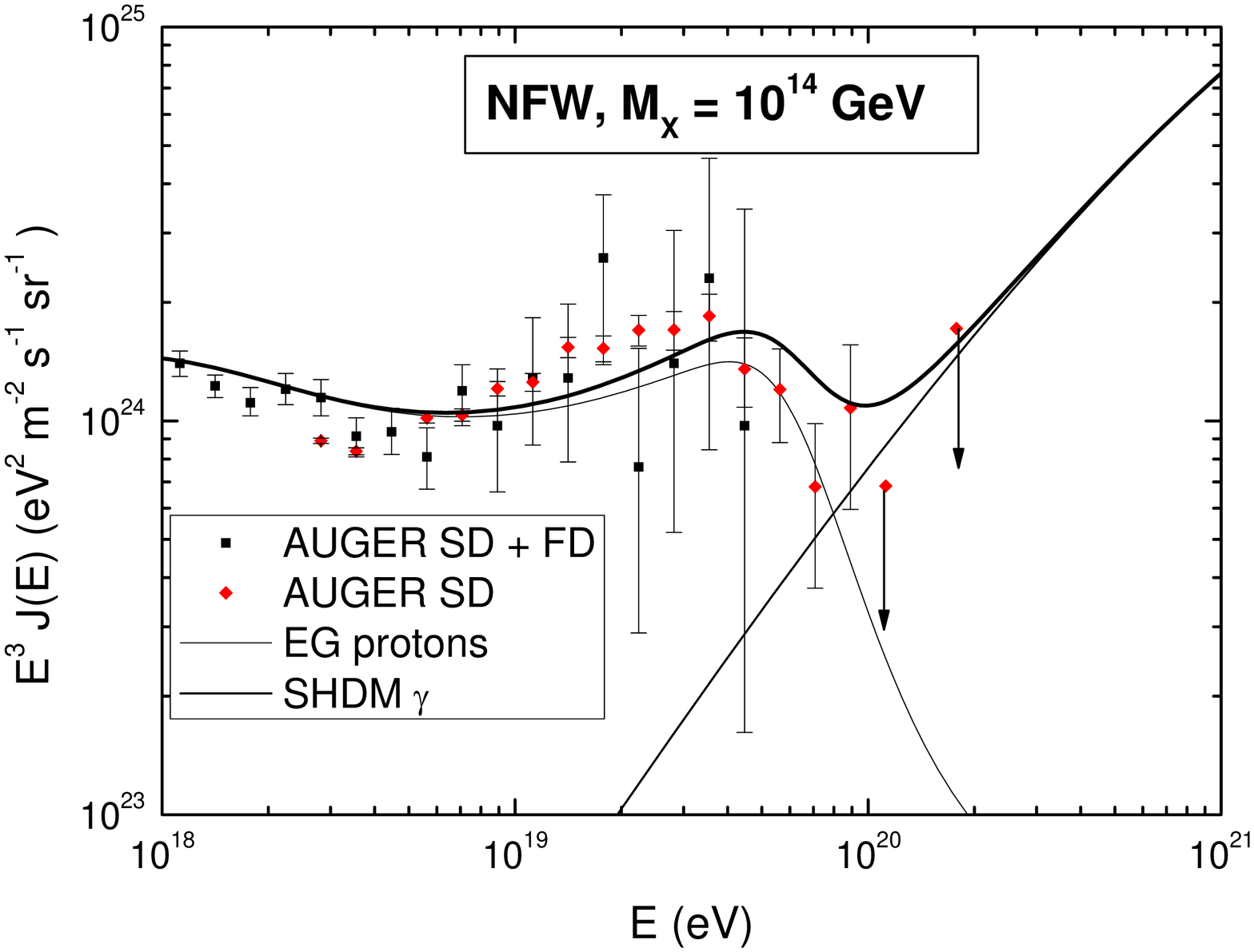}
\caption{The calculated spectrum of UHECRs from SHDM (dotted curve) in comparison with the Auger data ($M_X=10^{13}$ GeV left panel, $M_X=10^{14}$ GeV right panel). The continuos curve gives the
spectrum of extragalactic protons (assuming the Dip model \cite{Dip}). The sum of these two spectra is shown by thick continuos curve.}
\label{fig2}
\end{figure}

As it follows from equation  (\ref{flux1})  the flux from SHDM depends on two basic parameters: the X-particle life-time $\tau_X$ and mass $M_X$. While the value of $M_X$ is directly connected with the production mechanism of SHDM, its life-time can be regarded as a free parameter that should be fixed in order to fit the observed UHECR events. The most remarkable creation mechanism for SHDM is gravitational production through time variable gravitational fields during the inflationary phase of the universe \cite{grav}. In this case the mass $M_X$ fixes the present abundance of X-particles $\Omega_{SHDM}$. Assuming that SHDM constitute a substantial fraction of the DM, one has $\Omega_{SHDM}\simeq \Omega_{DM}$ and to provide $\Omega_{SHDM}=0.27$, according to the WMAP observations \cite{WMAP}, the mass of the X-particles must be in the range $10^{13}\div 10^{14}$ GeV.  Fixing the X-particle mass $M_X=10^{13}, 10^{14}$ GeV, in figure \ref{fig1} we plot the total UHECR flux (\ref{total1}) integrated over the AGASA field of view with the proper AGASA exposure. Fitting the AGASA excess we obtain the life time value $\tau_X\simeq 10^{20}$ y. The determination of the X-particle life time depends on the assumptions about the DM density profile with lower life-time in the case of the more concentrated Moore density profile (figure \ref{fig1} refers to the NFW density profile). In figure \ref{fig2} we have repeated the same computation of figure \ref{fig1} but in the case of the Auger detector. In this case the observed flux \cite{AugerFlux} is almost consistent with the GZK suppression, hence the SHDM contribution to the flux is {\it subdominant} with a typical life-time of the order of $\tau_X\simeq 10^{21}$ y.  

\begin{figure}[ht]
\includegraphics[width=0.52\textwidth]{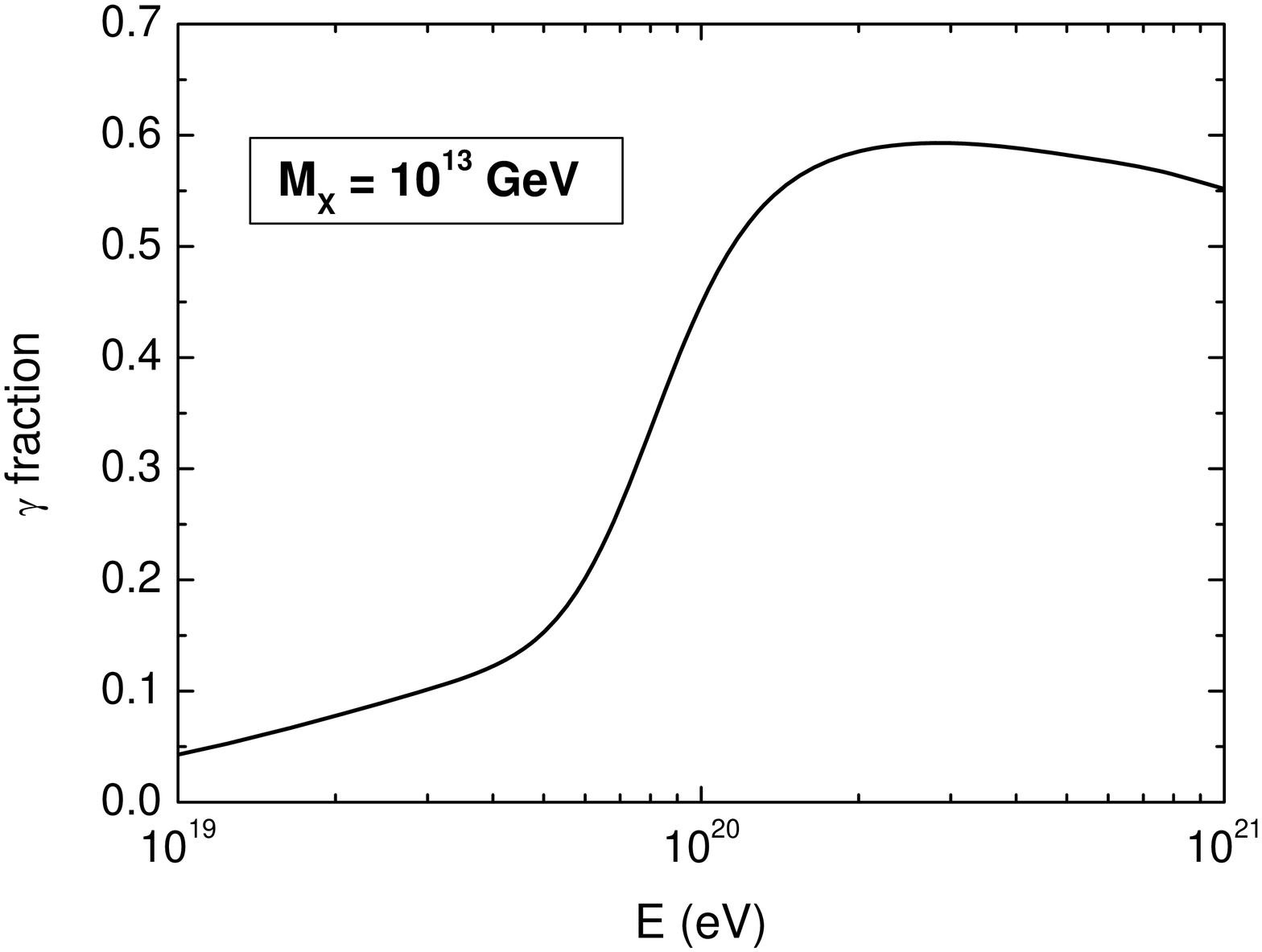}
\includegraphics[width=0.52\textwidth]{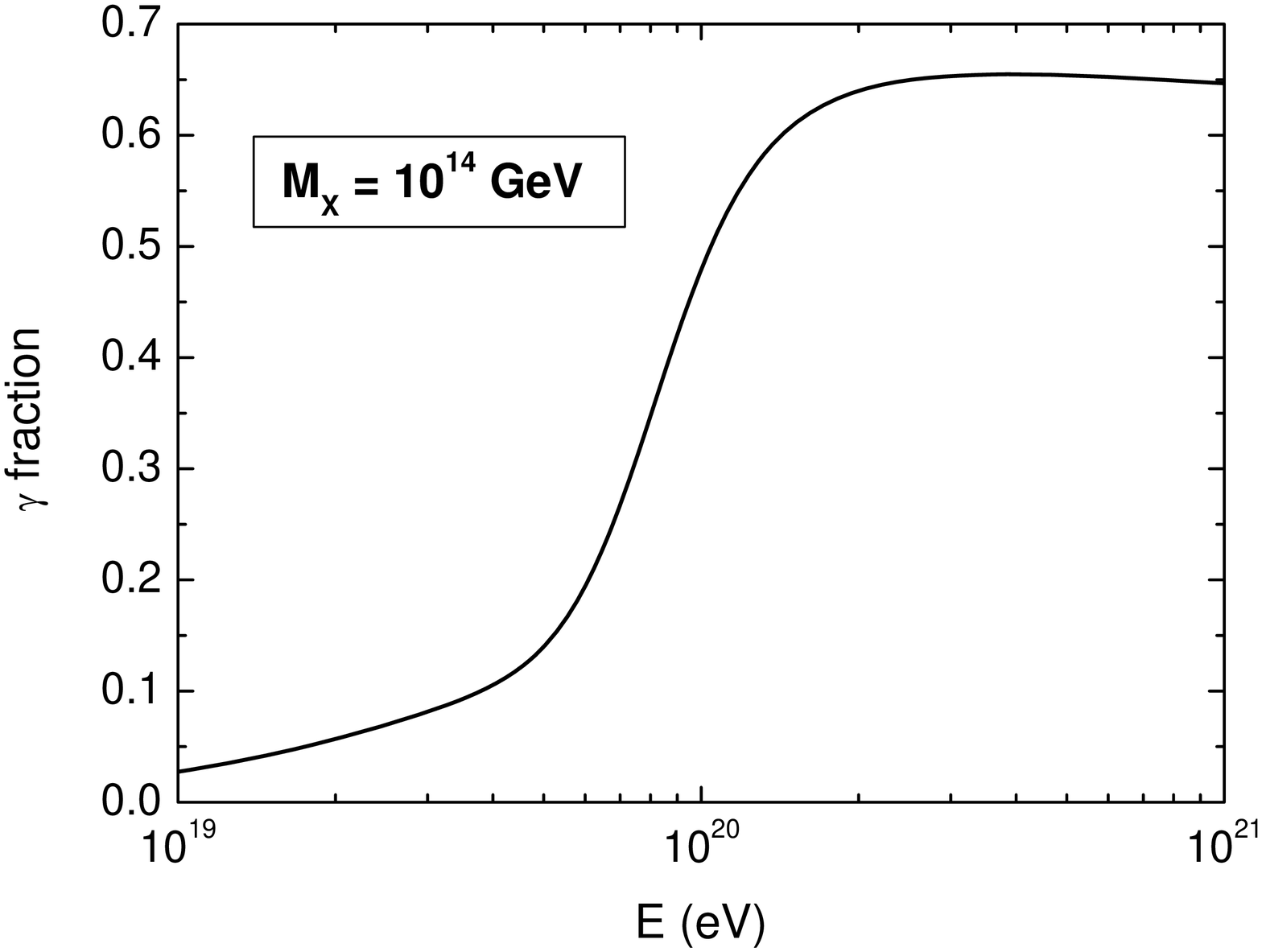}
\caption{Fraction of galactic SHDM photons in the total UHECR flux. The fluxes are computed in the case of the Auger detector assuming an NFW density profile and fixing $M_X=10^{13}$ GeV (left panel) and $M_X=10^{14}$ GeV (right panel).}
\label{fig3}
\end{figure}

We shall now discuss the expected UHECR chemical composition in the framework of SHDM production. In figure \ref{fig3} we have plotted the fraction of photons in the total flux computed under the same assumptions of figure \ref{fig2} (Auger observations, in the two cases $M_X=10^{13}$ GeV, left panel, and $M_X=10^{14}$ GeV, right panel). The result obtained, with a $2\div 3 \%$ photon content at $10^{19}$  is consistent with the Auger result that fixes the photon limit at $10^{19}$ eV to about $2\%$ \cite{AugerGamma,SemikozICRC07}, the same result holds at $10^{20}$ eV with an observed photon limit of $40\%$ in the Auger data and an expected photon fraction, as follows from figure \ref{fig3}, of the same order. As in the case of the  flux computation from the chemical composition analysis, we can conclude that, although disfavored, the SHDM hypothesis is not ruled out by the available observations.  

Let us conclude this section discussing the extragalactic contribution to the flux due to SHDM. The dominant photon component of this flux is absorbed by the intergalactic medium, it is completely converted in electromagnetic cascades in a few Mpc, see the discussion in \cite{BBV}. At the same time, the charged component (protons) suffers the interaction with CMB giving rise, at the highest energies, to the photo-pion production process. These interactions deplete the high energy flux of the charged extragalactic SHDM contribution, making such flux vanishingly small if compared with the corresponding astrophysical and galactic fluxes. For this reason we have neglected any extragalactic contribution to the SHDM flux. 

\section{Anisotropy}
\label{sec:ani}

The anisotropy in the direction of the GC is another important prediction of the SHDM model for the production of UHECR. In particular, being the SHDM distributed in our galactic halo with a spherical symmetry around the GC, a pure dipole deviation from isotropy might be expected with the dipole vector pointing toward the GC itself.

Independently of the top-down scenario, an anisotropy in the arrival directions of UHECR is also expected due to the cosmological Compton-Getting effect \cite{Sommers,Kachser}. At energies $E>10^{18}$ eV the UHECR flux is dominated by sources at cosmological distances, the movement of the Galaxy with respect to the CMB induces an anisotropy at 0.6\% level, in any case smaller than the signals we will discuss here.

As already anticipated in the introduction, we are mainly interested in the anisotropy expected at the Auger observatory. To compute it we introduce the relative acceptance $\omega$ of the detector, following \cite{Sommers}. The number of the detected CR events are indeed distributed in the sky depending on both the real celestial anisotropy and the detector relative exposure $\omega$.
The quantity $\omega$ can be computed in terrestrial equatorial coordinates assuming that the detector has a full-time operation, that is, a uniform exposure in right ascension $\alpha$. For a detector at latitude $\lambda$, fully efficient for particles arriving with a zenith angle $\theta<\theta_m$ (in the case of Auger,  $\lambda\simeq -35^\circ$ and $\theta_m=60^\circ$ for $E>3~10^{18}$ eV \cite{Augeracc}), then it is possible to deduce the dependence of the relative acceptance on the declination $\delta$:
\begin{equation}
\omega=\omega_0 \left [ \cos(\lambda)\cos(\delta)\sin(\alpha_m)+\alpha_m \sin(\lambda) \sin(\delta) \right ]
\label{exposure}
\end{equation}
where $\alpha_m$ is given by 
\begin{equation}
\alpha_m = \left\{ \begin{array}{ll}
\ 0 & \mbox{if $\xi > 1$}\\
\ \pi & \mbox{if $\xi < -1$}\\
\ cos^{-1}(\xi) & \mbox{otherwise}
 \end{array} \right.
\end{equation}
$$\xi \equiv \frac{cos(\theta_m)-sin(\lambda)sin(\delta)}{cos(\lambda)cos(\delta)}$$
and the constant $\omega_0$ is determined by the integrated acceptance of the observatory: 
$$2\pi \int_{\delta_{min}}^{\delta_{max}} \omega(\delta) \cos(\delta) d\delta= \frac{A_I}{S_A}~. $$
where $S_A$ is the surface covered by the observatory, $A_I=S_A\pi\sin^2(\theta_m)$ is its integrated acceptance and $(\delta_{min}=a_0-\theta_m,~\delta_{max}=a_0+\theta_m)$ is the observatory field of view in declination. For the fully completed Auger detector, $S_A=3000$ km$^2$  and $A_I\simeq 7000$~km$^2$~sr for $\theta_m=60^\circ$.

Taking into account the Auger observatory acceptance function, the expected UHECR flux can then be computed as a function of the equatorial terrestrial coordinates $(\alpha,\delta)$ 
\begin{equation}
J_{UHECR}(E,\alpha,\delta)=J_{EG}(E)\omega(\delta) + J_{SHDM}(E,\alpha,\delta)\omega(\delta)
\label{total2}
\end{equation}
where, as in equation (\ref{total1}), $J_{EG}$ is the extragalactic contribution (protons in the Dip model) and $J_{SHDM}$ is the galactic contribution from SHDM \footnote{The relation between the angle $\vartheta$, defined in equation (\ref{flux1}), and the galactic coordinates $(b,l)$ is $\cos(\vartheta)=\cos(b)\cos(l)$ and transforming to terrestrial equatorial coordinates ($\alpha,\delta$) one obtains $J_{SHDM}(E,\alpha,\delta)$ in equation (\ref{total2}).}. The SHDM contribution to the flux is mainly composed by photons with a small admixture of protons (not larger than $30\%$). At the lowest energies $E \sim 3\times 10^{18}$ eV protons are deflected by the galactic magnetic field and do not contribute to anisotropy. To take into account this effect and to bracket the possible anisotropy we have considered two extreme cases: (i) averaging, at all energies, the SHDM proton contribution over the sky and (ii) not averaging this contribution. From the point of view of the computed anisotropy the accuracy of these two approaches depends on energy, the case (i) is most reliable at low energy ($E\lsim 3\times 10^{19}$ eV) while case (ii) gives a better description at the highest energies ($E\gsim 3\times 10^{19}$ eV).

Using equation (\ref{total2}) it is possible to determine the number of events at energy $E\ge E_0$ expected to be collected during a time $T_0$. In order to apply an harmonic analysis, we integrate the events over $\delta$, in 24 bins in $\alpha$ (i.e., $\Delta \alpha=15^\circ$):
\begin{equation}
N_{\alpha_i}(\ge E_0)=S_A T_0 \int_{\alpha_i}^{\alpha_i+\Delta\alpha} d\alpha \int_{E_0} dE 
\int_{\delta_{min}}^{\delta_{max}} J_{UHECR}(E,\alpha,\delta) \cos(\delta) d\delta
\label{numb}
\end{equation}

\begin{figure}[ht]
\includegraphics[width=0.52\textwidth]{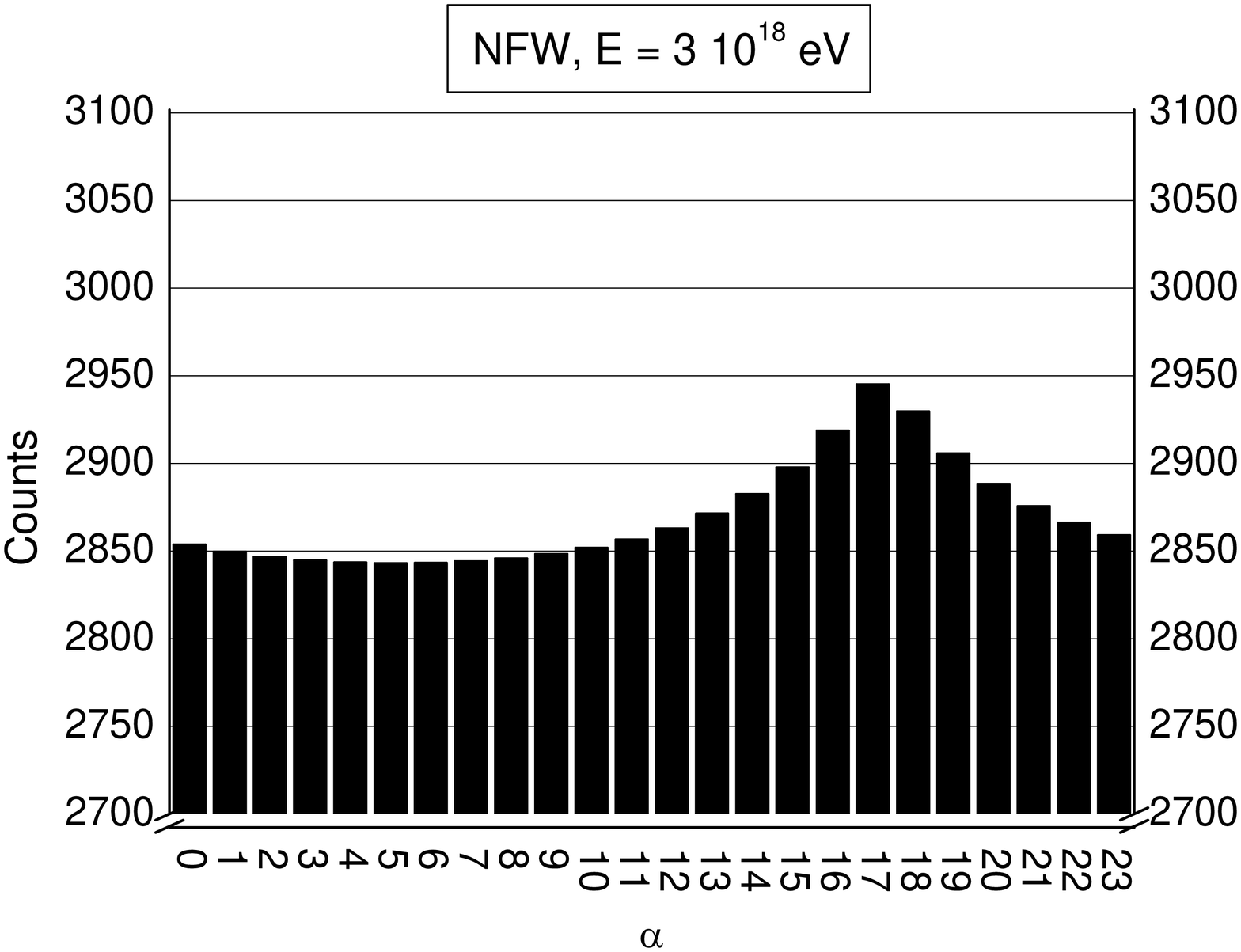}
\includegraphics[width=0.52\textwidth]{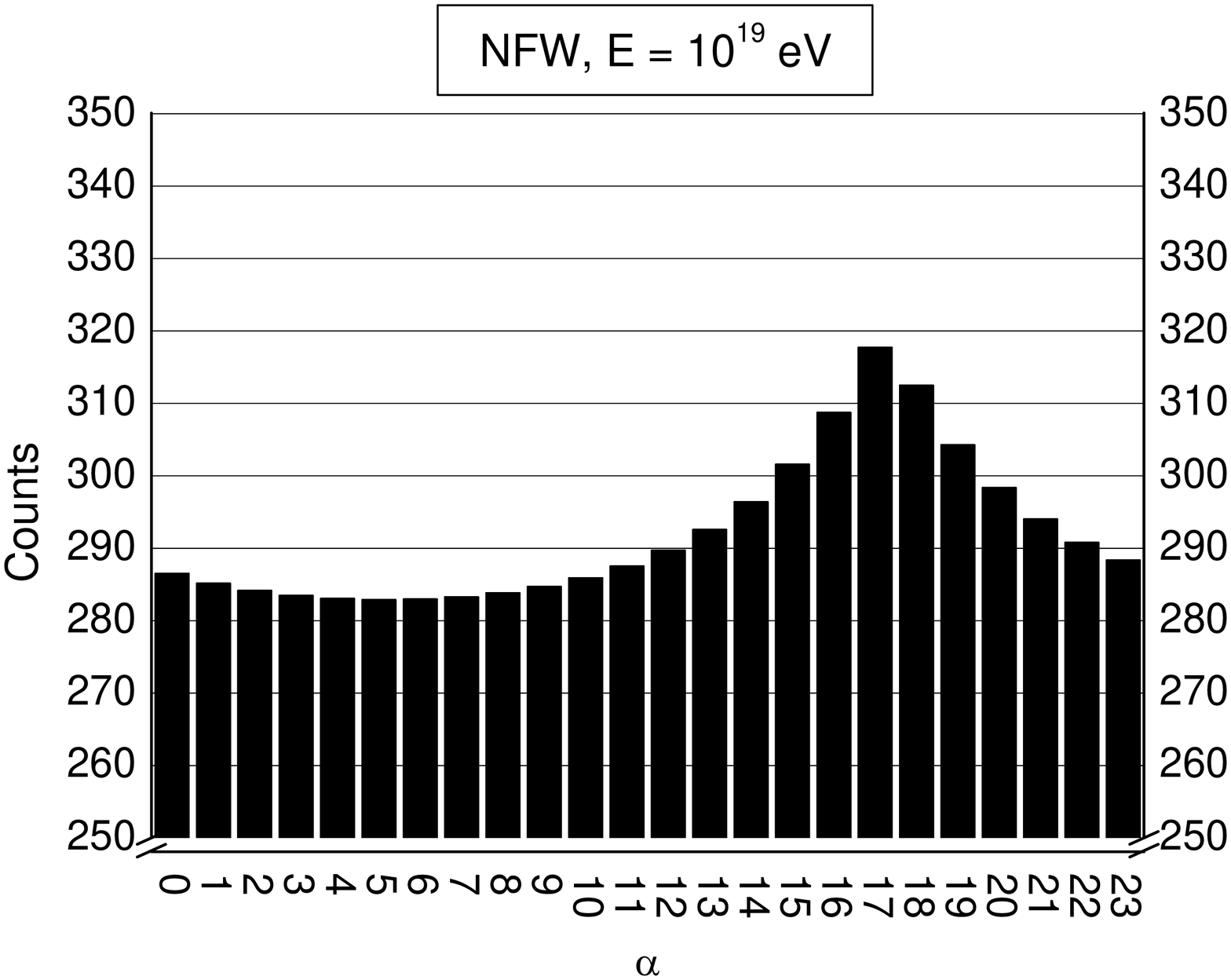}
\includegraphics[width=0.52\textwidth]{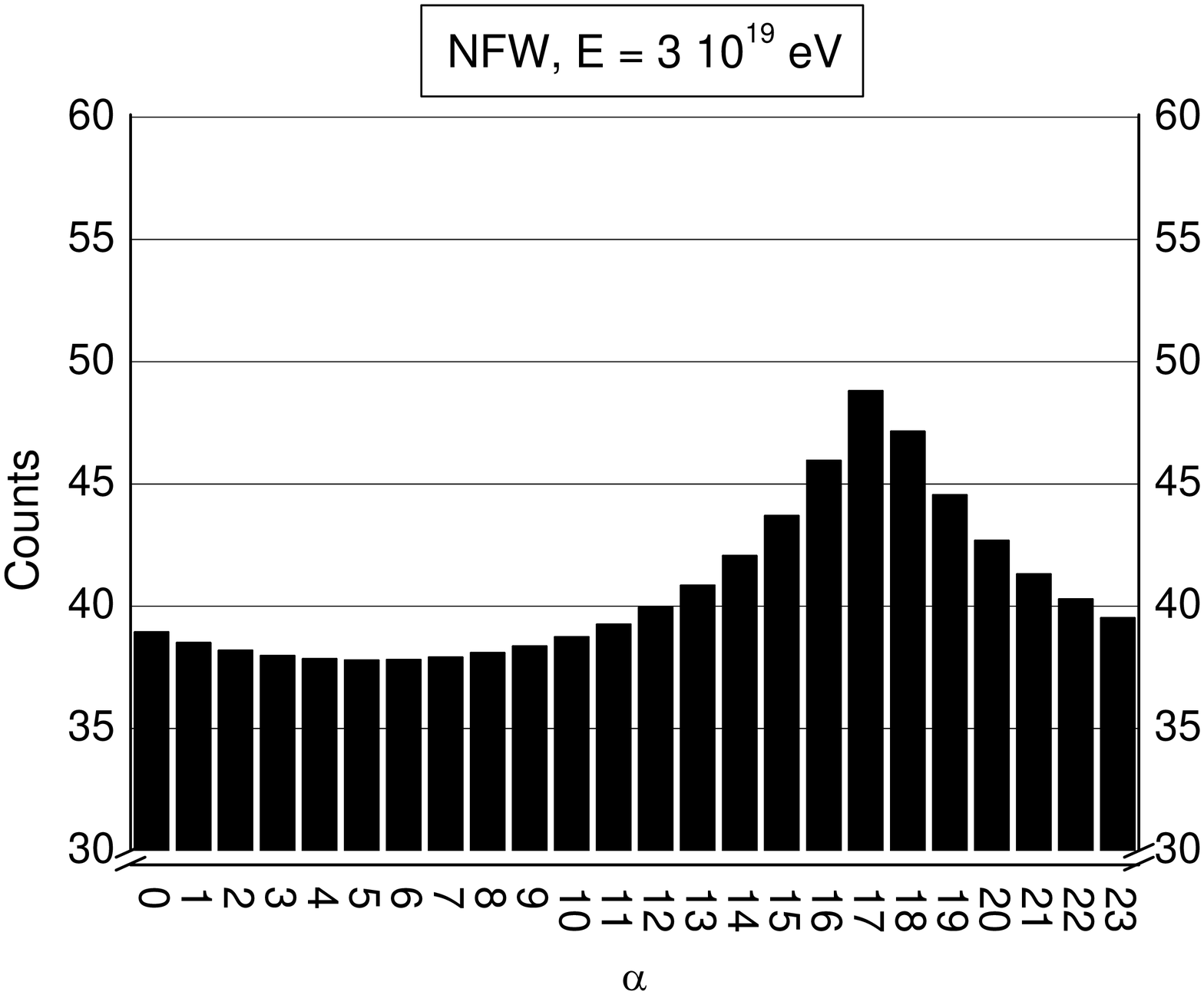}
\includegraphics[width=0.52\textwidth]{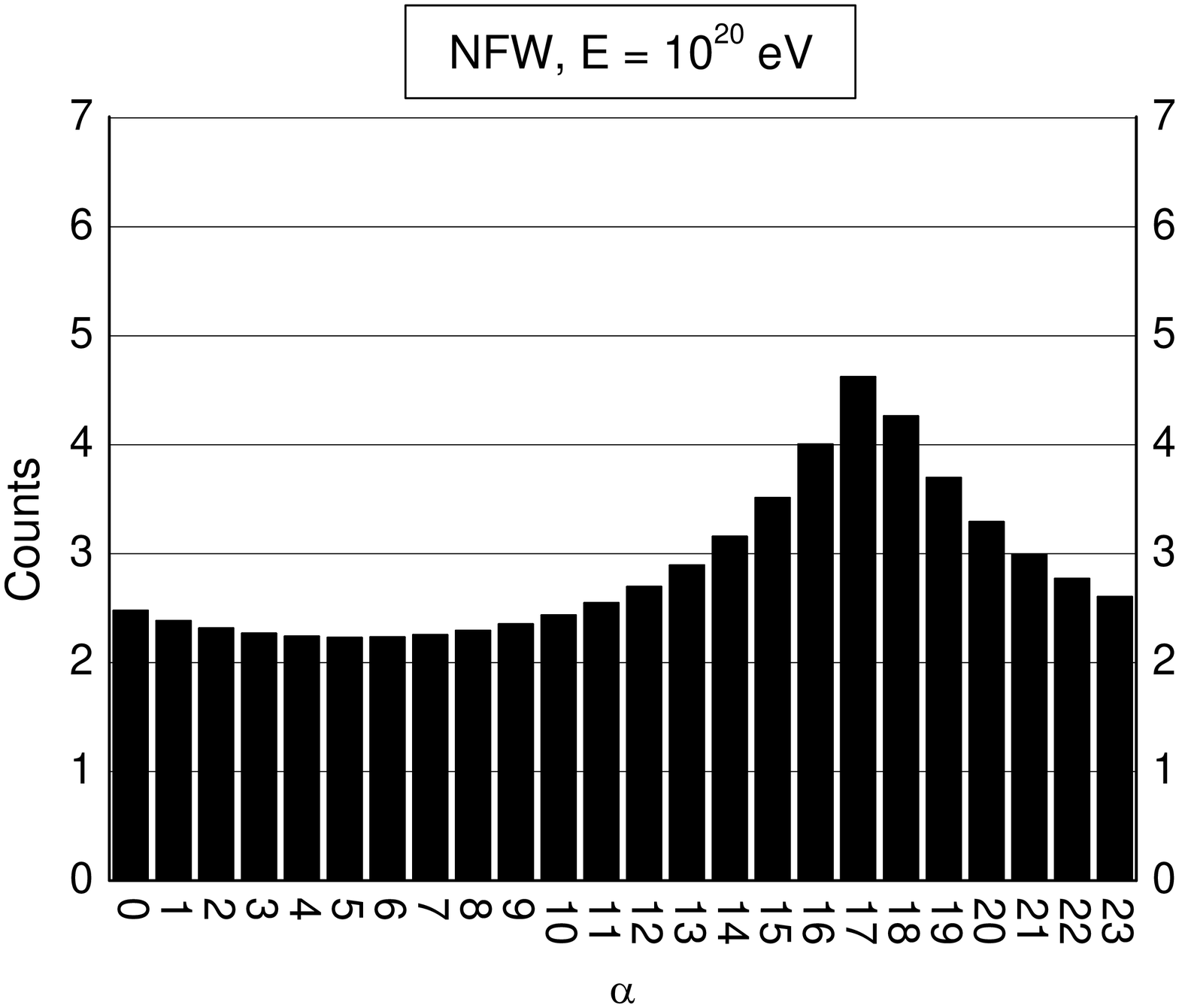}
\caption{The expected UHECR events at the Auger observatory in 5 years data taking as function of right ascension. The SHDM contribution is obtained in the case of NFW density profile, assuming a
SHDM mass of $M_X=10^{13}$ eV and averaging, at all energies, the SHDM proton component over the sky.}
\label{fig4}
\end{figure} 

\begin{figure}[ht]
\includegraphics[width=0.52\textwidth]{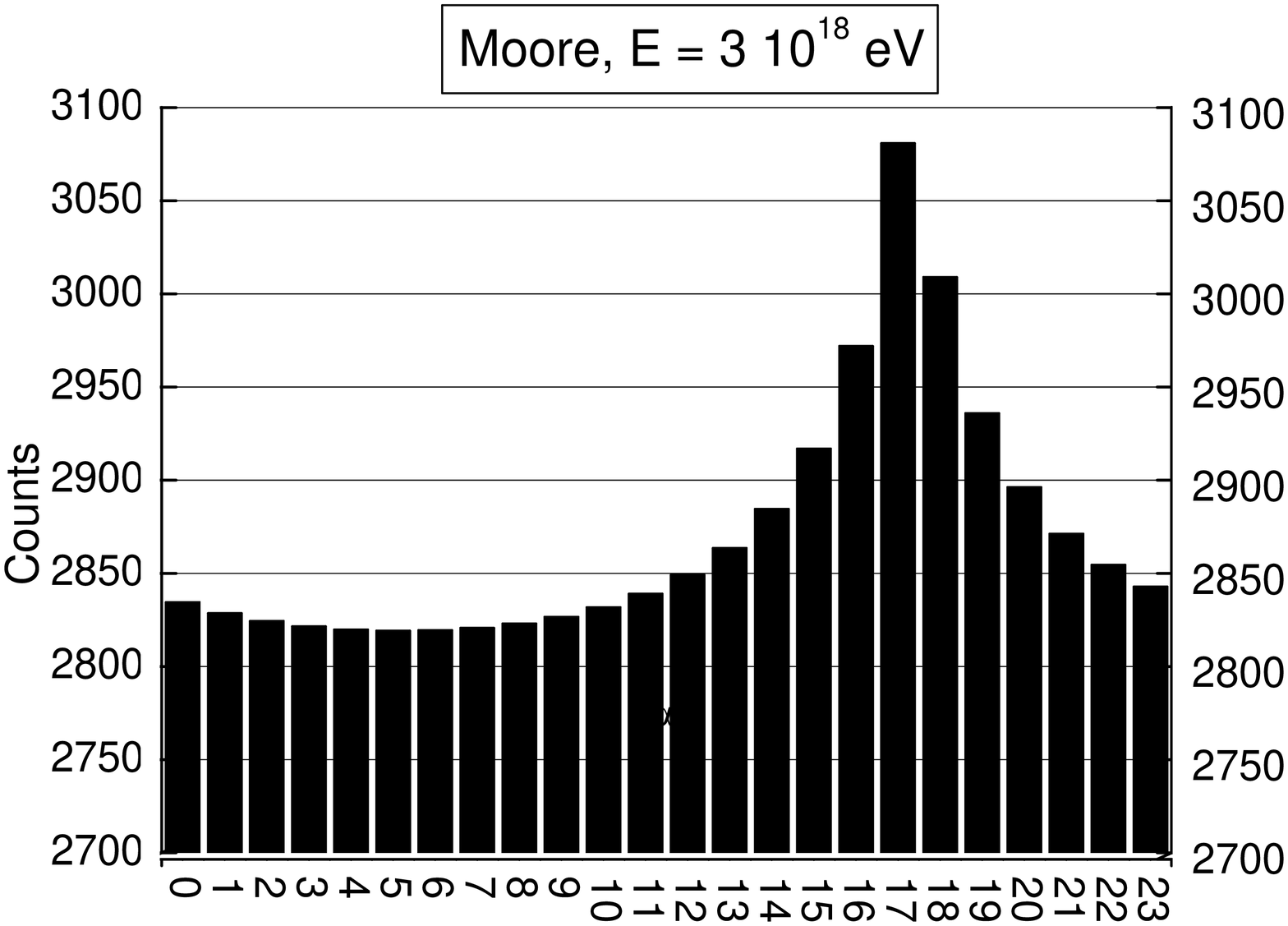}
\includegraphics[width=0.52\textwidth]{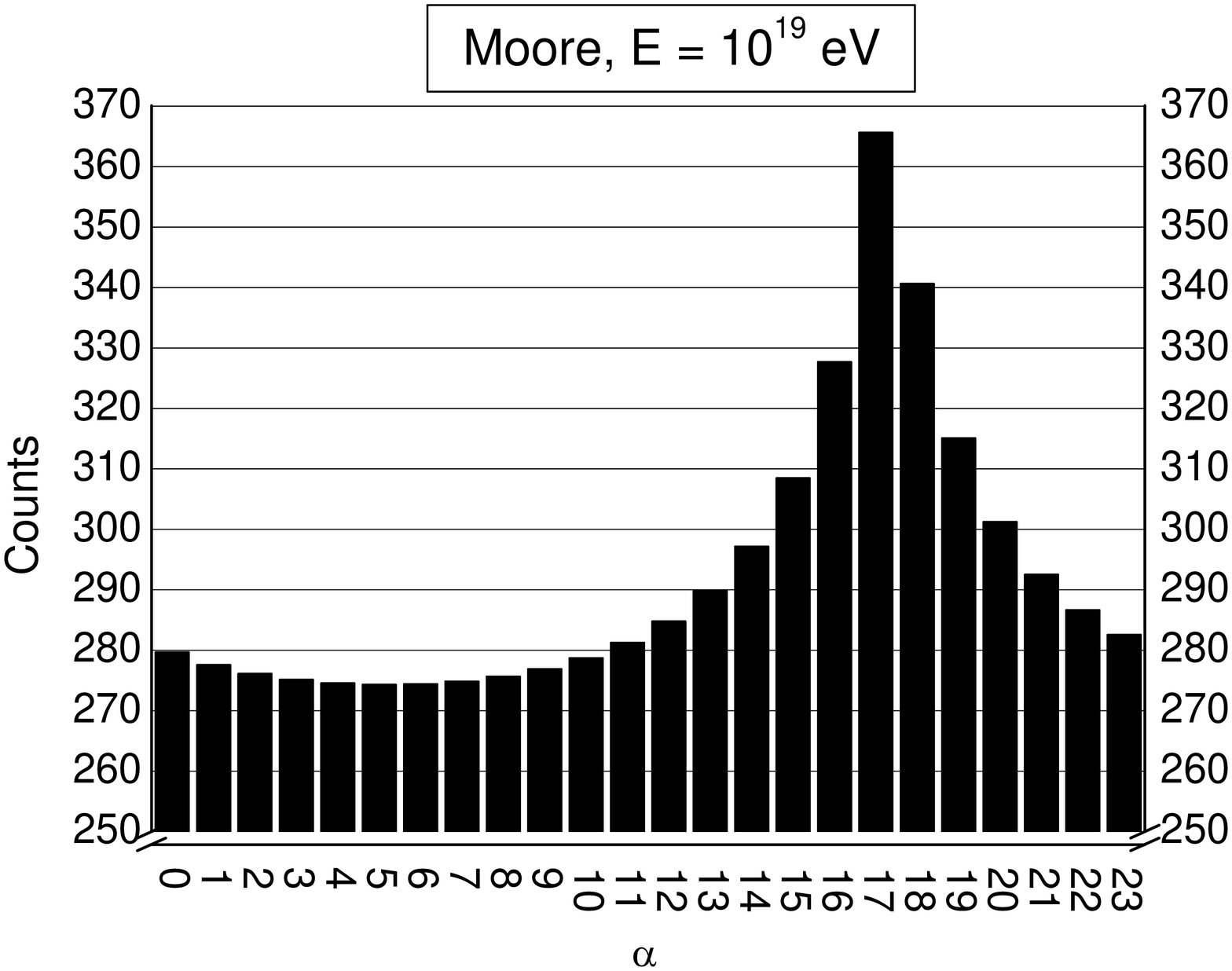}
\includegraphics[width=0.52\textwidth]{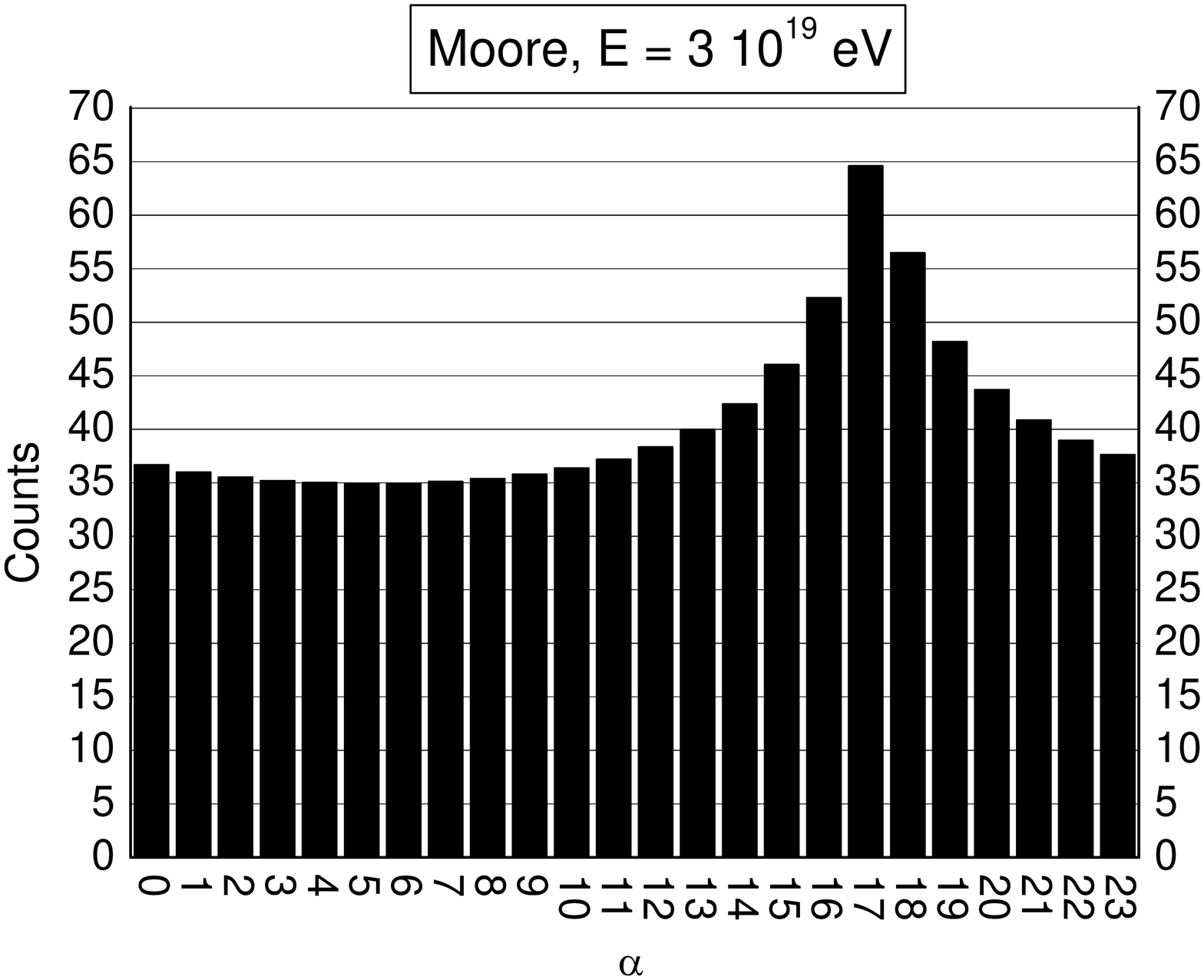}
\includegraphics[width=0.52\textwidth]{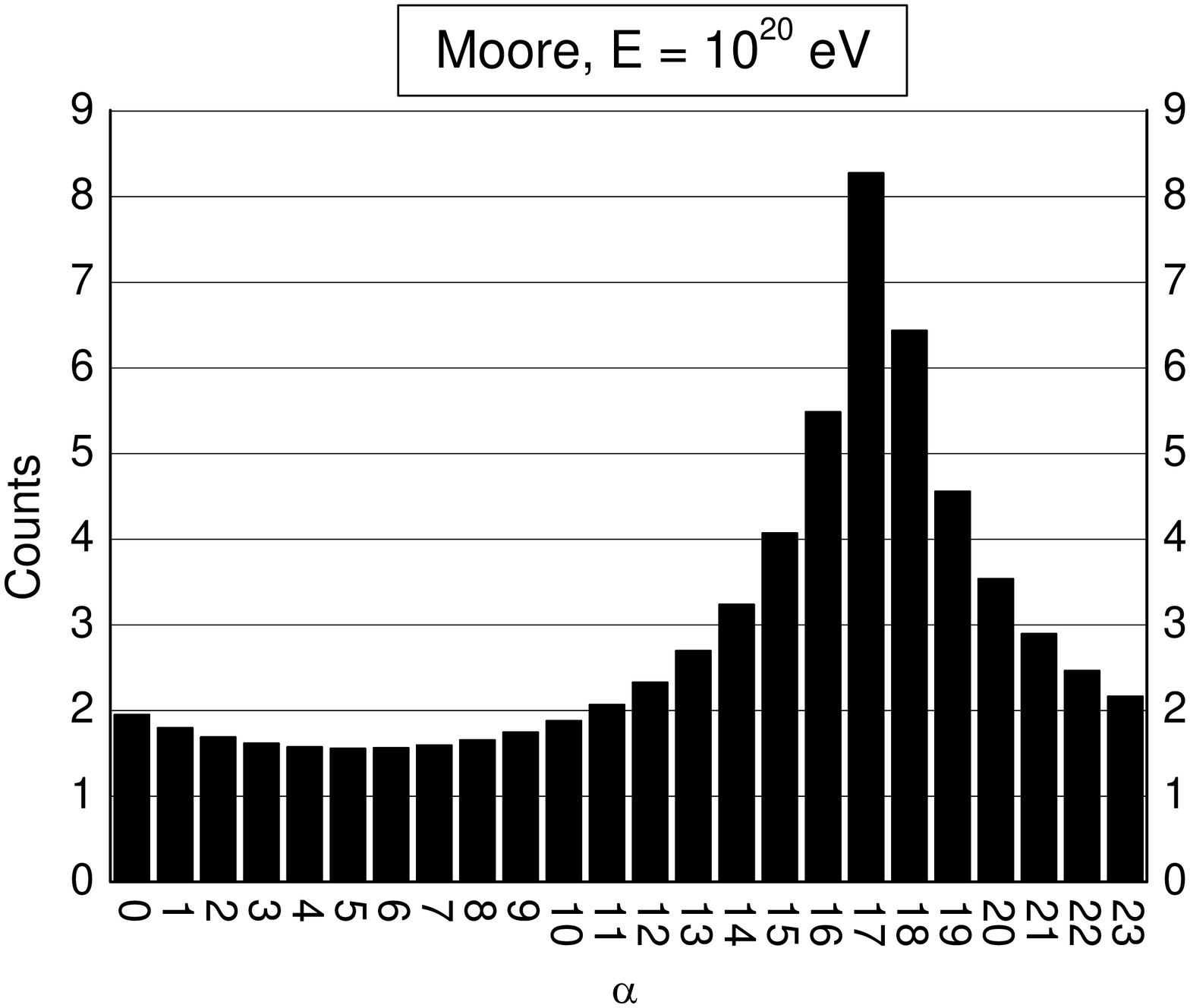}
\caption{The expected UHECR events at the Auger observatory in 5 years data taking as function of  right ascension. The SHDM contribution is obtained in the case of Moore density profile, assuming a SHDM mass of $M_X=10^{13}$ eV and keeping, at all energies, the angular dependence of the SHDM proton component.}
\label{fig5}
\end{figure}

In figure \ref{fig4} and \ref{fig5} we show, in the case of NFW and More respectively, the right ascension distribution of the expected events at the Auger observatory after $T_0=5$ yr period of data taking: the four panels in each figure correspond to four different energies, $E_0=3\times 10^{18}$,~$10^{19}$,~$3\times 10^{19}$,~$10^{20}$~eV as labeled. In figure \ref{fig4} we have averaged the SHDM proton component 
over the sky while in figure \ref{fig5} we kept the angular dependence of the SHDM proton flux. These two figures can be regarded as the two possible extreme cases of lowest (figure \ref{fig4}) and highest anisotropy (figure \ref{fig5}). 

To study the anisotropy we apply the Rayleigh's formalism (see e.g. \cite{Linsley}) to derive the amplitude and phase of the first harmonic in right ascension. Within this method the amplitude $r$ and phase $\phi$ are given by: $r=\sqrt{a^2+b^2}$ and $\phi=atan(\frac{b}{a})$, where $a=\frac{2}{N}\sum_{i=1}^Ncos(\alpha_i)$, $b=\frac{2}{N}\sum_{i=1}^Nsin(\alpha_i)$, and N=24 in our case. In table \ref{tab1} we show the amplitude of the first harmonic in the case of the two density profiles  and in the four different energy ranges. In table \ref{tab1} we have also quoted the two cases of minimum and maximum anisotropy obtained respectively averaging and not averaging the SHDM proton contribution over the sky. We do not list the phase because in all cases it always result to be $\phi=17.7$ hr (the galactic center coordinates being $\alpha_{GC}=$ 17h 42m 30s $\delta_{GC}=-28^\circ$ $55'$ $18''$).

The quantity of interest is then the number of events, $N_{evs}$, necessary to detect such amplitude and phase with a defined significance, $S$. Since the uncertainty on the amplitude is given by $\sqrt {\frac{2}{N_{evs}}}$, we derive that $N(S)=\frac{2S^2}{r^2}$. 
Thus to get a measurement of the expected anisotropies at 4 s.d. level, the number of required events is listed in the same table respectively for the two density profiles and in the case of minimum and maximum expected anisotropy. Such number of events roughly correspond to a maximum of 7 years and a minimum of 2 years of data taking in the Auger observatory. The Moore case indeed corresponds to a more concentrated distribution of X-particles and thus gives a stronger signal from the GC, that will be detectable with less statistics. It is moreover important to stress that, in both density cases, already in the first energy bin under study, i.e., $E_0=3\times 10^{18}$ eV, the anisotropy is at the level of few $\%$ and can be detectable in a few years by Auger.  

As already discussed the SHDM induced anisotropy is connected with the DM distribution in the galaxy, thus it is expected to appear on a typical scale that corresponds to the fiducial radius $R_s$ of this distribution as defined in equation (\ref{profile}). The fiducial radius $R_s$ as a typical value of $45$ Kpc \cite{BM} and it corresponds to an expected anisotropy on a typical scale of $60^\circ$ as follows from figures \ref{fig4} and \ref{fig5}. This is a large scale anisotropy if compared with the anisotropies already studied by Auger in the GC direction on a typical scale of a few degrees \cite{AugerSmallAni}. 

The result obtained is of particular importance because it shows the possibility of testing the SHDM hypothesis already at low energies where the Auger observatory has a larger statistic. Moreover, in the present evaluation we have bracketed the possible anisotropy taking into account the two cases of a SHDM proton component averaged and not averaged over the sky.  
\begin{table}[H]
\vspace{-0.3cm}
\begin{center}
\begin{tabular}{|c|cc|cc|}
\hline\hline
Energy (EeV) & $r$ (NFW-min) & N(4 s.d.) & $r$ (Moore-min) & N(4 s.d.) \\
\hline
$>3$ & 0.015 & 150000 & 0.018 & 100000 \\[0.1cm]
$>10$ & 0.045  & 16000 & 0.060 & 9000 \\[0.1cm]
$>30$ &  0.100 & 3200 & 0.140 & 1600 \\[0.1cm]
$>100$& 0.300 & 350 & 0.440 & 150 \\[0.1cm]
\hline
Energy (EeV) & $r$ (NFW-max) & N(4 s.d.) & $r$ (Moore-max) & N(4 s.d.) \\[0.1cm]
\hline
$>3$ & 0.019 & 90000 & 0.027 & 45000 \\[0.1cm]
$>10$ & 0.065 & 7500 & 0.092 & 4000 \\[0.1cm]
$>30$ & 0.150 & 1400 & 0.215 & 700 \\[0.1cm]
$>100$& 0.480 & 150 & 0.690 & 70 \\[0.1cm]
\hline\hline
\end{tabular}
\end{center}
\vspace{1.pc} \caption{Amplitude of first harmonic in right ascension, and number of events required to a detection at 4 standard deviations. Upper panel shows the case with minimum anisotropy (with isotropization of SHDM protons), lower panel shows the case 
of maximum anisotropy (with no isotropization of SHDM protons).}
\label{tab1}
\end{table}
The first case gives a reliable anisotropy at low energy ($E\lsim 10^{19}$ eV) where the effect of the galactic magnetic field is important; the second case is more reliable at the highest energies ($E\gsim 3 \times 10^{19}$ eV) where the galactic magnetic field has no effect on the UHECR proton propagation.

\section{Conclusions}
\label{sec:concl}
In the present paper we have investigated the hypothesis of SHDM as a source of a {\it subdominant} component in the observed UHECR flux. The SHDM hypothesis is a viable candidate for the DM in the universe. Super heavy particles are naturally produced in time varying gravitational fields during the inflation era of the universe. In this context the observed DM density fixes the expected mass of SHDM in the range ($10^{13}$~GeV,~$10^{14}$~GeV). Moreover, SHDM can be quasi-stable, due to a super-weak discrete gauge symmetry breaking, with a life time exceeding the age of the universe.  These characteristics of SHDM are of particular importance from the point of view of UHECR because they make SHDM a natural candidate for the production of ultra high energy particles. In order to explain the observed UHECR spectrum with SHDM the only free parameter of the model is the SHDM particles life-time $\tau_X$. 

Super heavy particles are accumulated in the halo of our galaxy with a typical overdensity of $2.1\times 10^{5}$ thus the corresponding UHECRs can be unabsorbed by CMB not showing the GZK steepening. Among UHECR observations only the AGASA data, with an excess of events violating the GZK suppression,
can be used to fix a value of the SHDM life-time; from the HiRes, Yakutsk and Auger data, almost compatible with the GZK cut-off, follows only a lower bound of the SHDM life time. In any case these flux observations do not exclude the SHDM hypothesis and its role in the UHECR production can still be investigated as a {\it subdominant} contribution to the flux. 

The basic signatures of the SHDM model for the UHECR production are a flat injection spectrum $\propto E^{-1.9}$, a photon (and neutrino) dominated flux and an anisotropy due to the off-center position of the earth in the galaxy. In the present paper we have analyzed both the chemical composition and anisotropy expected at the Auger observatory, our basic findings can be summarized as follows.

\begin{itemize}
\item{The observed Auger spectrum  is still compatible with a {\it subdominant} contribution from SHDM with a typical life-time of $\tau_X \gsim 10^{21}$ y.}
\item{The chemical composition observed by Auger, with a photon limit around $2\%$ at $10^{19}$ eV, is still compatible with the chemical composition expected in the SHDM model, that shows a photon content of the spectrum at the level of few $\%$ at $10^{19}$ eV.}
\item{The anisotropy expected by SHDM can be detected by the Auger observatory and, even at the lowest energies, can be measured with a good accuracy.}
\end{itemize}

The anisotropy signal obtained in the present evaluation is of particular importance because it shows that already at low energy the SHDM hypothesis can be tested. At the lowest Auger energies  $E\gsim 10^{18}$ eV the charged component of the UHECR flux is isotropized by the galactic magnetic field while the photon dominated flux from SHDM does not. This is an important result connecting, at low energy, any observed anisotropy in the flux to the SHDM component. In the present paper we have performed a detailed computation of the expected anisotropy signal at Auger, showing how this signal is within reach of the Auger capabilities. 

Finally, at the highest energies, the anisotropy amplitude computed here should be regarded as an upper limit (see the discussion in section \ref{sec:ani}). Therefore, the lack of detection by Auger in 5 or 7 years of full operation would not discard the SHDM hypothesis and, probably, a very large aperture new generation experiment will be needed to settle down the question about UHECR produced by the decay of SHDM.  

\section*{Acknowledgments}
We are grateful to V. Berezinsky for joint work on the present paper, we also thank P. Blasi for valuable discussions and P. Ghia for her collaboration in anisotropy computations. The present work is partly funded by ASI through the contract ASI-INAF I/088/06/0 on High Energy Astrophysics.

\end{document}